\documentclass[aps,twocolumn,prl,tightenlines,floatfix,superscriptaddress]{revtex4-1}

\usepackage[breaklinks=true]{hyperref}
\usepackage{graphicx}
\usepackage[english]{babel}
\usepackage{amsmath}
\usepackage{amssymb}
\usepackage{times}

\begin{document}

\newcommand{\D}{\mathrm{D}}
\newcommand{\p}{\partial}
\newcommand{\Tr}{\mathrm{Tr}}
\renewcommand{\d}{\mathrm{d}}
\newcommand{\Ek}{E_\mathbf{k}}
\newcommand{\xik}{\xi_\mathbf{k}}
\newcommand{\sumk}{\sum_\mathbf{k}}

\title{Unusual destruction and enhancement  of superfluidity of atomic
  Fermi gases by population imbalance in a one-dimensional optical
  lattice}

\author{Qijin Chen} \email[Email: ]{qchen@uchicago.edu}
 \affiliation{Shanghai Branch, National Laboratory for Physical Sciences at Microscale and Department of Modern Physics, University
  of Science and Technology of China, Shanghai 201315, China}
\affiliation{Department of Physics and Zhejiang Institute of Modern
  Physics, Zhejiang University, Hangzhou, Zhejiang 310027, China}
\affiliation{Synergetic Innovation Center of Quantum Information and
  Quantum Physics, Hefei, Anhui 230026, China}

\author{Jibiao Wang}
\email[Email: ]{wangjibiao@gmail.com}
\affiliation{Laboratory of Quantum Engineering and Quantum Metrology, School of Physics and Astronomy, Sun Yat-Sen University (Zhuhai Campus), Zhuhai, Guangdong 519082, China}

\author{Lin Sun}
\affiliation{Department of Physics and Zhejiang Institute of Modern
  Physics, Zhejiang University, Hangzhou, Zhejiang 310027, China}

\author{Yi Yu }
\affiliation{College of Chemical Engineering, Zhejiang University of Technology, Hangzhou, Zhejiang 310014, China}

\date{\today}

\begin{abstract}

  We study the superfluid behavior of a population imbalanced
  ultracold atomic Fermi gases with a short range attractive
  interaction in a one-dimensional (1D) optical lattice, using a
  pairing fluctuation theory. We show that, besides widespread
  pseudogap phenomena and intermediate temperature superfluidity, the
  superfluid phase is readily destroyed except in a limited region of
  the parameter space. We find a new mechanism for pair hopping,
  assisted by the excessive majority fermions, in the presence of
  continuum-lattice mixing, which leads to an \emph{unusual} constant
  BEC asymptote for $T_c$ that is independent of pairing strength. In
  result, on the BEC side of unitarity, superfluidity, when it exists,
  may be strongly \emph{enhanced} by population imbalance.
\end{abstract}

\maketitle

Ultracold atomic Fermi gases have been an ideal system for quantum
simulation and quantum engineering, due to their multiple tunable
parameters \cite{Review,Bloch_RMP}. Using a Feshbach resonance, one
can vary the effective pairing strength from the weak coupling BCS
limit to strong pairing BEC limit. As another widely explored
parameter \cite{ZSSK06,Rice1,Stability,Sheehy_RPP,YD05,PWY05,FGLW05},
population imbalance leads to Fermi surface mismatch and thus renders
pairing more difficult, causing suppressed superfluid transition
temperature $T_c$ or complete destruction of superfluidity at high
population imbalances \cite{Stability}. Among other tunable parameters
is the geometry of the system; one can put the Fermi gas in an optical
lattice, such as a one-dimensional (1D) optical lattice (OL), which we
shall explore here. Unlike the widely studied 3D continuum or 3D
lattice (for which each lattice site contains 1 or 2 fermions), 1DOL
is distinct in that it is a lattice-continuum mixed system with each
lattice ``site'' now containing many fermions. Such a system has not
been properly studied in the literature, with and without a population
imbalance.

Population imbalance $p$ has been widely known to suppress or destroy
superfluidity. Indeed, in a 3D homogeneous system, superfluidity is
completely destroyed at $T=0$ in the unitary and BCS regimes
\cite{Chien06,Stability,chen07prb}, leaving only possible intermediate
temperature superfluids (ITSF). Nonetheless, in the BEC regime, stable
superfluid exists even with very high $p$, and all minority fermions
are paired up. This has been naturally understood as a consequence of
vanishing Pauli blocking effect in the deep BEC regime, where the
distribution of the constituent fermions in a Cooper pair spreads out
over the entire momentum space and thus pairs can happily coexist with
excessive fermions.

In this Letter, we show that the pairing and superfluid behavior of a
Fermi gas, when subject to a short-range attractive interaction
($U<0$) in 1DOL with $p\ne 0$, is very different due to the
lattice-continuum mixing. We find that superfluidity may be readily
destroyed by population imbalance, except for a very restricted
parameter range (away from small $t$, large $p$, and large lattice
constant $d$), where population imbalance gives rise to an extra
mechanism for pair hopping.  When a BEC superfluid does exist, this
leads to an \emph{unusual constant} BEC asymptote for $T_c$, and then
can substantially increase $T_c$ on the BEC side of unitarity as
compared to the balanced case, in which $T_c$ decreases with
interaction strength.  In addition, not all minority fermions are
paired up in the BEC limit.  We demonstrate that these unusual
behaviors are associated with the mixing of a 2D continuum plane and a
discrete lattice dimension, which leads to a constant ratio of
$\Delta^2/\mu$, unlike in a 3D continuum or 3D lattice. This mixing
enables enhanced pair hopping processes assisted by excessive
fermions.

There have seemingly been many theoretical studies on Fermi gases in
optical lattices
\cite{Chen1,Demler2,Bloch_NP,Michael05PRL,Cazalilla05PRL,OrsoPRL95,Torma06NJP,Chien08PRA_RC,Stringari_RMP}. However,
most studies have used the chemical potentials and magnetization as
control variables and are thus limited to the weak and intermediate
pairing strength regimes.  In a 3D attractive Hubbard model with
\cite{Micnas14AP} and without \cite{Chen1} population imbalance,
superfluid in the deep BEC regime exists only at low fillings.
On the other hand, 1DOL of $^6$Li has been realized
experimentally with and without population imbalance
\cite{OngThomasPRL114,ThomasPRL120}. However, its phase diagram is yet
to be explored \cite{noteonOL}.
We \emph{emphasize} that 1DOL is fundamentally a 3D system, albeit
anisotropic. It \emph{cannot} be compared with a genuine 2D or 1D
lattice, which has usually no more than 2 fermions per site, and does
\emph{not} support \emph{true} long-range order as we study here.

Here we use a previously developed pairing fluctuation theory
\cite{ChenPRL98,Review,chen07prb}. It goes beyond the BCS mean-field
treatment by self-consistently including finite momentum pairing in
the self energy, which thus contains two parts,
$\Sigma(K) = \Sigma_{sc}(K) + \Sigma_{pg}(K)$, where
$\Sigma_{sc}(K) = -\Delta_{sc}^2G_0(-K)$ and
$\Sigma_{pg}(K) = \sum_Q t_{pg}(Q)G_0(Q-K)$, corresponding to the
contributions of the Cooper pair condensate and finite momentum pairs,
respectively.  We shall follow the notations of Ref.~\cite{ChenPRL98},
such that $\hbar=k_B=1$, and four momenta
$K\equiv (\omega_n, \mathbf{k})$, $Q\equiv (\Omega_l, \mathbf{q})$,
$\sum_Q\equiv T\sum_l\sum_\mathbf{q}$, etc. Here $G_0(K)$ is the
non-interacting Green's function, $t_{pg}(Q)$ the $T$ matrix,
$\Delta_{sc}$ the order parameter, and $\omega_n$ ($\Omega_l$) the odd
(even) Matsubara frequency. The finite momentum pairing 
directly leads to the presence of a pseudogap when it becomes
strong.
This theory has been applied to 3D homogeneous and trapped Fermi gases
\cite{Chen14Review,M2S2009,chen07prb}, as well as on a 3D or quasi-2D
lattice \cite{Chen1,Chien08PRA_RC}, and has been used by other groups
\cite{Micnas14AP,Torma2,PhysRevA.74.031604,PhysRevA.87.053616}.

Now we adapt this theory for 1DOL by modifying the noninteracting
atomic dispersion into
$\xi_{\mathbf{k}\sigma}=\epsilon_\mathbf{k}-\mu_\sigma \equiv
\mathbf{k}_\parallel^2/2m + 2t[1-\cos (k_zd)]-\mu_\sigma$, where
$\mathbf{k}_\parallel=(k_x,k_y)$ is the in-plane momentum, and
$\mu_\sigma$ the chemical potential for spin
$\sigma=\uparrow,\downarrow$. This one-band lattice dispersion is
justified when the band gap in the $z$ direction is tuned to be much
greater than the Fermi energy in the $xy$ plane.  The derivation of
our self-consistent equations is otherwise the same, so that we shall
present the result directly, with an emphasis on the \emph{unusual}
new findings caused by population imbalance and the lattice-continuum
mixing.

In the superfluid phase, we  define the pseudogap via
$\Delta_{pg}^2 = -\sum_Q t_{pg}(Q)$, so that the total gap $\Delta$ is
given by $\Delta^2 = \Delta_{sc}^2 + \Delta_{pg}^2$, which leads to
the self energy $\Sigma_\sigma(K) \approx -\Delta^2 G_{0,\bar{\sigma}}(-K)$, and the
full Green's function
\begin{equation}
 G_{\sigma}(K)=\frac{u_{\textbf{k}}^{2}}{i\omega_{n}-E_{\textbf{k}\sigma}}+\frac{v_{\textbf{k}}^{2}}{i\omega_{n}+E_{\textbf{k}\bar{\sigma}}}\,,
\end{equation}
where $u_{\textbf{k}}^{2}=(1+\xi_{\textbf{k}}/E_{\textbf{k}})/2$,
$v_{\textbf{k}}^{2}=(1-\xi_{\textbf{k}}/E_{\textbf{k}})/2$,
$E_{\textbf{k}\uparrow}=E_{\textbf{k}}-h$,
$E_{\textbf{k}\downarrow}=E_{\textbf{k}}+h$, and
$E_{\textbf{k}}=\sqrt{\xi_{\textbf{k}}^{2}+\Delta^{2}}$,
$\xi_{\textbf{k}}=\epsilon_{\textbf{k}}-\mu$,
$\mu=(\mu_{\uparrow}+\mu_{\downarrow})/2$,
$h=(\mu_{\uparrow}-\mu_{\downarrow})/2$.  
Then we have the number equations,
\begin{eqnarray}
  n&=&\sum_{\textbf{k}}\Big[\Big(1-\frac{\xi_{\textbf{k}}}{E_{\textbf{k}}}\Big)
 +2\bar{f}(E_{\textbf{k}})\frac{\xi_{\textbf{k}}}{E_{\textbf{k}}}\Big]\,, 
\label{eq:LOFF_neqa}\\
 p n&=&\sum_{\textbf{k}}\Big[f(E_{\textbf{k}\uparrow})-f(E_{\textbf{k}\downarrow})\Big]\,,
\label{eq:LOFF_neqb}
\end{eqnarray}
where  $p=(n_\uparrow-n_\downarrow)/n$,
$\bar{f}(x)=[f(x+h)+f(x-h)]/2$, and $f(x)=1/(e^{x/T}+1)$.
%
We have the following gap equation with pair chemical potential
$\mu_{p}=0$ in the superfluid phase, 
\begin{equation}
  \frac{m}{4\pi a}=\sum_{\textbf{k}}\Big[\frac{1}{2\epsilon_{\textbf{k}}}-\frac{1-2\bar{f}(E_{\textbf{k}})}{2E_{\textbf{k}}}\Big]+a_{0}\mu_{p}\,,
  \label{eq:gap}
\end{equation}
where the interaction $U$ has been replaced by the $s$-wave scattering
length $a$ via
$U^{-1}=m/4\pi a-\sum_{\textbf{k}}1/2\epsilon_{\textbf{k}}$. Here a
finite $\mu_{p}$ extends this equation into the non-superfluid phase.
We caution that the parameter $a$ does \emph{not} necessarily yield
the experimentally measured scattering length, which is better
reflected by an effective scattering length $a_\text{eff}$ such that
$1/a_\text{eff}=\sqrt{2mt}d/a$. (See Supplementary Secs.~I, II and
Fig.~S1).  The coefficient $a_0$ is determined via Taylor expanding
$t_{pg}^{-1}(Q)$ on the real frequency axis,
$t_{pg}^{-1}(\Omega,\textbf{q})\approx
a_{1}\Omega^{2}+a_{0}(\Omega-\Omega_{\textbf{q}}+\mu_{p})$, with
$\Omega_{\textbf{q}}=B_{\parallel}\textbf{q}_{\parallel}^{2} +
2t_{B}[1-\cos(q_{z}d)]$.  Here $B_{\parallel}=1/2M_\parallel$, with
$M_\parallel$ being the effective pair mass in the $xy$-plane, and
$t_{B}$ is the effective pair hopping integral.
Then we have the pseudogap equation
\begin{equation}
  a_0\Delta_{pg}^{2}=\sum_{\textbf{q}}\frac{b(\tilde{\Omega}_{\textbf{q}})}{\sqrt{1+4\dfrac{a_{1}}{a_{0}}(\Omega_{\textbf{q}}-\mu_{p})}}\,,
  \label{eq:PG}
\end{equation}
with $b(x)=1/(e^{x/T}-1)$ 
and pair dispersion
$\tilde{\Omega}_{\textbf{q}}=\{\sqrt{a_{0}^{2}[1+4a_{1}(\Omega_{\textbf{q}}-\mu_{p})/a_{0}]}-a_{0}\}/2a_{1}$,
which reduces to
$\tilde{\Omega}_{\textbf{q}}=\Omega_{\textbf{q}}-\mu_p$ when
$a_1/a_0 \ll 1$, e.g., in the BEC regime.

Equations (\ref{eq:LOFF_neqa})-(\ref{eq:PG}) form a closed set of
self-consistent equations, which will be solved for ($\mu_{\uparrow}$,
$\mu_{\downarrow}$, $\Delta_{pg}$, $T_c$) with $\Delta_{sc}=0$, and
for ($\mu_{\uparrow}$, $\mu_{\downarrow}$, $\Delta$, $\Delta_{pg}$) in
the superfluid phase.  For our numerics, we consider $p>0$, and define
Fermi momentum $k_{F}=(3\pi^{2}n)^{1/3}$ and Fermi energy
$E_{F}\equiv k_{B}T_{F}=\hbar^{2}k_{F}^{2}/2m$ \cite{noteonEf}.

The asymptotic solution in the BEC limit, $\mu \rightarrow -\infty$,
can be obtained analytically fully for $p=0$ or partially for
$p>0$. For $p=0$, $\mu_\sigma = \mu$. However, for $p>0$, we have
$\mu_\uparrow > 0$ at $ T_c$ throughout the BCS-BEC crossover, and
$\mu_\downarrow = 2\mu -\mu_\uparrow$. This is self-consistently
justified by the solution that $T_c \ll \Delta \ll |\mu|$ in the BEC
regime. Therefore, in the BEC limit,
$f(\Ek^\downarrow) = f(\xik^\downarrow) = 0$ for all $p$, but
$f(\Ek^\uparrow) = f(\xik^\uparrow) = 0$ only for $p=0$.
From the number equations, we obtain
\begin{equation}
  \label{eq:muBECp}
  \mu =  -te^{d/a} + 2t + \frac{2\pi dn_\uparrow}{m}\,,
\end{equation}
dominated by the leading two-body term. The exponential dependence of
$\mu$ on $d/a$ results from the quasi-two
dimensionality 
since $|k_z| \le \pi/d$. In the $t\rightarrow 0$ 2D limit, one finds
$\frac{d}{a} \approx \ln\frac{|\mu|}{t}$, which diverges
logarithmically.  To leading order corrections in powers of $1/\mu$,
we have
\begin{eqnarray}
  \label{eq:nBECp}
  (1-p)n &=& -\frac{m\Delta^2}{4\pi \mu d} -\frac{np\Delta^2}{2\mu^2}, \quad \text{or}\\
  \Delta &=& \sqrt{\frac{4\pi|\mu|d(1-p)n}{m}} \left(1-\frac{\pi dnp}{\mu m} \right)\,.
  \label{eq:gapBECp}
\end{eqnarray}
At $T_c$, $\Delta_{pg} = \Delta$. Note that $\Delta^2/\mu$ approaches
a constant in the BEC limit, in contrast to its counterpart in a 3D
homogeneous case, where $\Delta\sim |\mu|^{1/4}$ so that
$\Delta^2/\mu\rightarrow 0$.

For $p=0$, one can easily obtain
\begin{equation}
B_\parallel = \frac{1}{4m}, \quad \mbox{and}\quad
t_B = \frac{t^2}{2|\mu|} \approx \frac{t}{2} e^{-d/a},
\label{eq:BtB}
\end{equation}
which yields
\begin{equation}
  T_c \approx \dfrac{\pi a n}{2m} = \dfrac{k_Fa}{3\pi}T_F
  \label{eq:p0Tc}
\end{equation}
in the BEC regime via the pseudogap equation (\ref{eq:PG}).

Now for $p>0$, one has to solve for $\mu_\uparrow$ and $T_c$
numerically, since $\mu_\uparrow > 0$. We have
$\Ek^\uparrow \approx \xik^\uparrow + \frac{4\pi d n_\downarrow}{m}$,
then Eq.~(\ref{eq:LOFF_neqb}) becomes
\begin{equation}
  pn = \sum_\mathbf{k} f\Big(\xik^\uparrow + \frac{4\pi d n_\downarrow}{m}\Big) .
 \label{eq:nupBEC}
\end{equation}

In the BEC regime, $T_c$ is controlled by the inverse $T$-matrix
expansion. The coefficients $a_0$ (and pair density $n_p$) and $a_1$
are given by
\begin{eqnarray}
 \label{eq:a0BECp}
  n_p = a_0 \Delta^2
&=& n_\downarrow -\sumk\bigg[f(\xik^\uparrow)-f\Big(\xik^\uparrow + \frac{4\pi d n_\downarrow}{m}\Big)\bigg],\hspace*{4ex}\\
a_1\Delta^2 &=& A  + \frac{n_\downarrow}{4|\mu|} (1 + B )\,.
\label{eq:a1BECp}
\end{eqnarray}
For the inverse pair mass, we have
\begin{equation}
  B_\parallel = \frac{1}{4m} + \delta B_\parallel\,,\quad
  t_B  = \frac{t^2}{n_p}\bigg[C
       + \frac{n_\downarrow}{2|\mu|}(1 -D )\bigg].
          \label{eq:BzBECp}
\end{equation}
Here $A$, $B$, $C$, $D$ and $\delta B_\parallel $ depend on
$\mu_\uparrow$ and $T$ only. They can be readily obtained via the
inverse $T$-matrix expansion (see Supplementary Sec.~III for concrete
expressions).
For $p=0$, $\mu_\uparrow \rightarrow -\infty$ so that
$f(\xik^\uparrow)$, as well as $A$, $B$, $C$, $D$ and
$\delta B_\parallel $ all vanish. Then we recover
$n_p = n_\downarrow = n/2$, $a_1\Delta^2 = -n/8\mu$ and
Eq.~(\ref{eq:BtB}).

Of \emph{paramount importance} is that population imbalance leads to these
extra terms in $a_0\Delta^2$, $a_1\Delta^2$, $B_\parallel$ and $t_B$,
which are associated with the  excessive majority fermions via
the Fermi functions.
Equation (\ref{eq:BzBECp}) suggests that the pair motion in the $z$
direction for $p\ne 0$ is now strongly enhanced by these fermions as
an extra pair hopping mechanism; a minority fermion may hop to the
next site by exchanging its majority partner to one that is already
there, which is guaranteed by the existence of a transverse continuum
dimension, since the ``site'' is actually a 2D plane.  Obviously, this
extra mechanism will be dominant over the usual virtual pair
unbinding-rebinding process \cite{NSR} in the BEC regime.  Indeed, the
pair hopping integral $t_B$ and thus $T_c$ approach constant BEC
asymptotes rather than decreasing with $1/k_Fa$.  Furthermore,
Eq.~(\ref{eq:a0BECp}) indicates that $n_p < n_\downarrow$ for
$p\ne 0$, namely, not all minority fermions are paired in the BEC
limit.

In the deep BEC regime, Eq.~(\ref{eq:muBECp}) determines $\mu$, and
Eq.~(\ref{eq:gapBECp}) yields the gap $\Delta$, for given
$1/k_Fa$.  Then $\mu_\uparrow$ and $T_c$ can be obtained via solving
Eq.~(\ref{eq:PG}) (with $\Delta_{pg}=\Delta$) along
with 
Eq.~(\ref{eq:nupBEC}), with the help of Eqs.~(\ref{eq:a0BECp}) and
(\ref{eq:a1BECp}) \cite{noteonequations}. Finally,
$\mu_\downarrow = 2\mu - \mu_\uparrow$.

\begin{figure}
  \includegraphics[clip,width=3.2in]{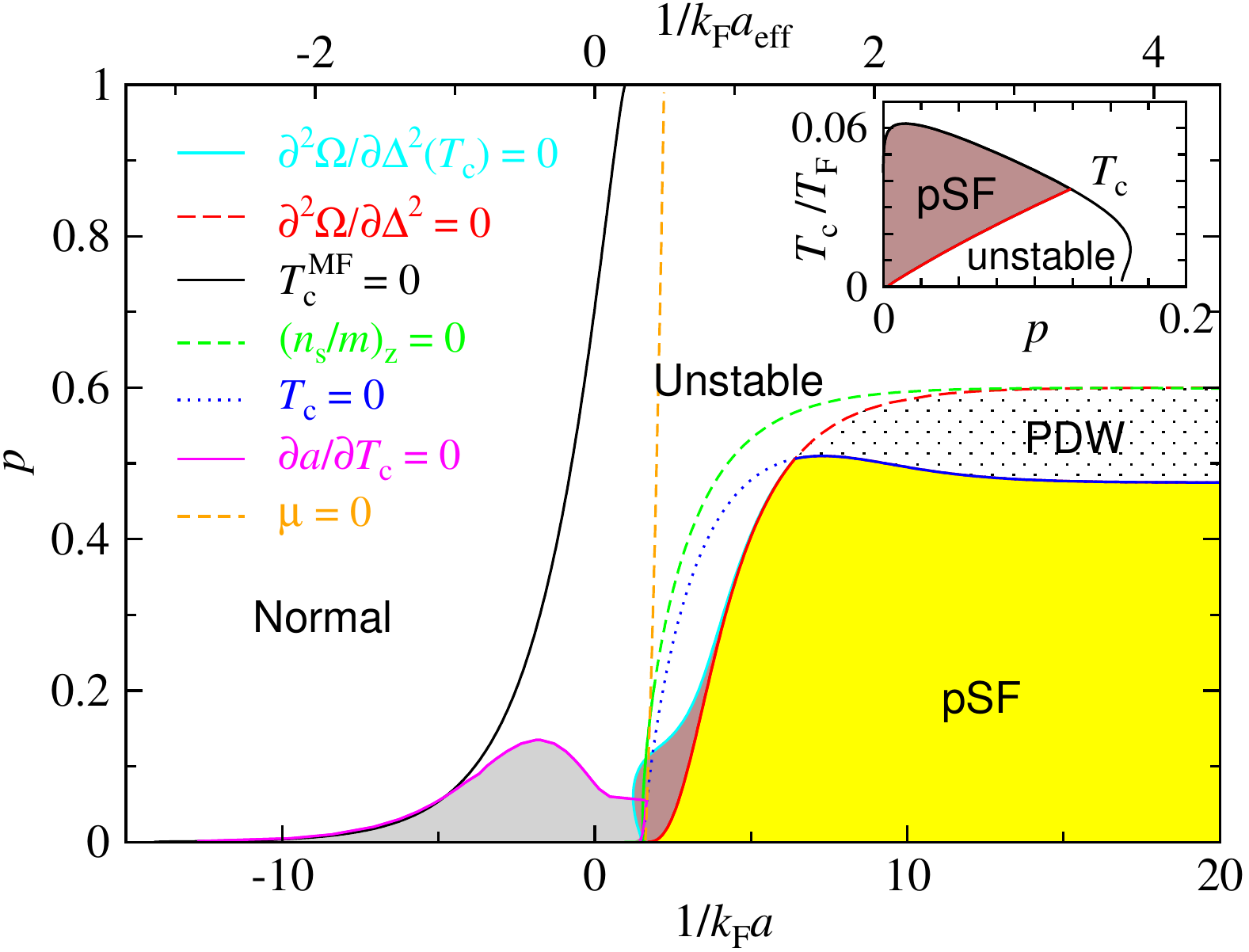}
  \caption{Zero temperature phase diagram in the $p$ -- $1/k_Fa$
    plane, for $t/E_F=0.2$ and $k_Fd = 0.5$. The orange dashed $\mu=0$
    line separates the fermionic and bosonic regimes. Here ``Normal''
    denotes normal Fermi gases. Besides the pSF (yellow shaded) and
    PDW (dotted region), and normal phases, the mean-field superfluid
    solution between the black (solid) and red (solid and dashed)
    curves is unstable at $T=0$.  The grey and brown areas admit
    ITSF. An example in the brown area is shown in the inset for
    $1/k_Fa=2$. The green dashed and blue dotted lines show less
    restrictive instability conditions, with $n_s/m$ being superfluid
    density. The top axis is labeled with $1/k_Fa_\text{eff}$.}
  \label{fig:T0phase}
\end{figure}

Shown in Fig.~\ref{fig:T0phase} the zero $T$ phase diagram for
$(t/E_F,k_Fd)= (0.2,0.5)$. These parameters allow a relatively large
polarized superfluid (pSF) phase (yellow shaded region) in the BEC
regime.  A normal phase lies to the left of the black solid line of
$T_c^\text{MF}=0$. The red line is given by the instability condition
$\partial^2 \Omega/\partial\Delta^2=0$ against phase separation,
following Refs.~\cite{Chien06,Stability}, where $\Omega$ is the
thermodynamic potential. The blue $T_c=0$ curve is determined by
$t_B=0$. A possible pair density wave (PDW) state emerges in the
dotted region where $t_B$ becomes negative \cite{ChePRA2016}.  The
rest space has an unstable mean-field solution of pSF at $T=0$.  The
grey and brown shaded regions allow ITSF; the former has a lower
$T_c$, whereas the latter does not but is unstable at low $T$, as
shown in the inset for $1/k_Fa = 2$. To compare with the 3D continuum
case \cite{Chien06}, we label the top axis with
$1/k_Fa_\text{eff}$. Apparently, the position of the Normal/Unstable
boundary is roughly the same, but stable pSF solution no longer exists
here for high $p\gtrsim 0.5$. Instead, a PDW phase emerges.

\begin{figure}
  \includegraphics[clip,width=3.2in]{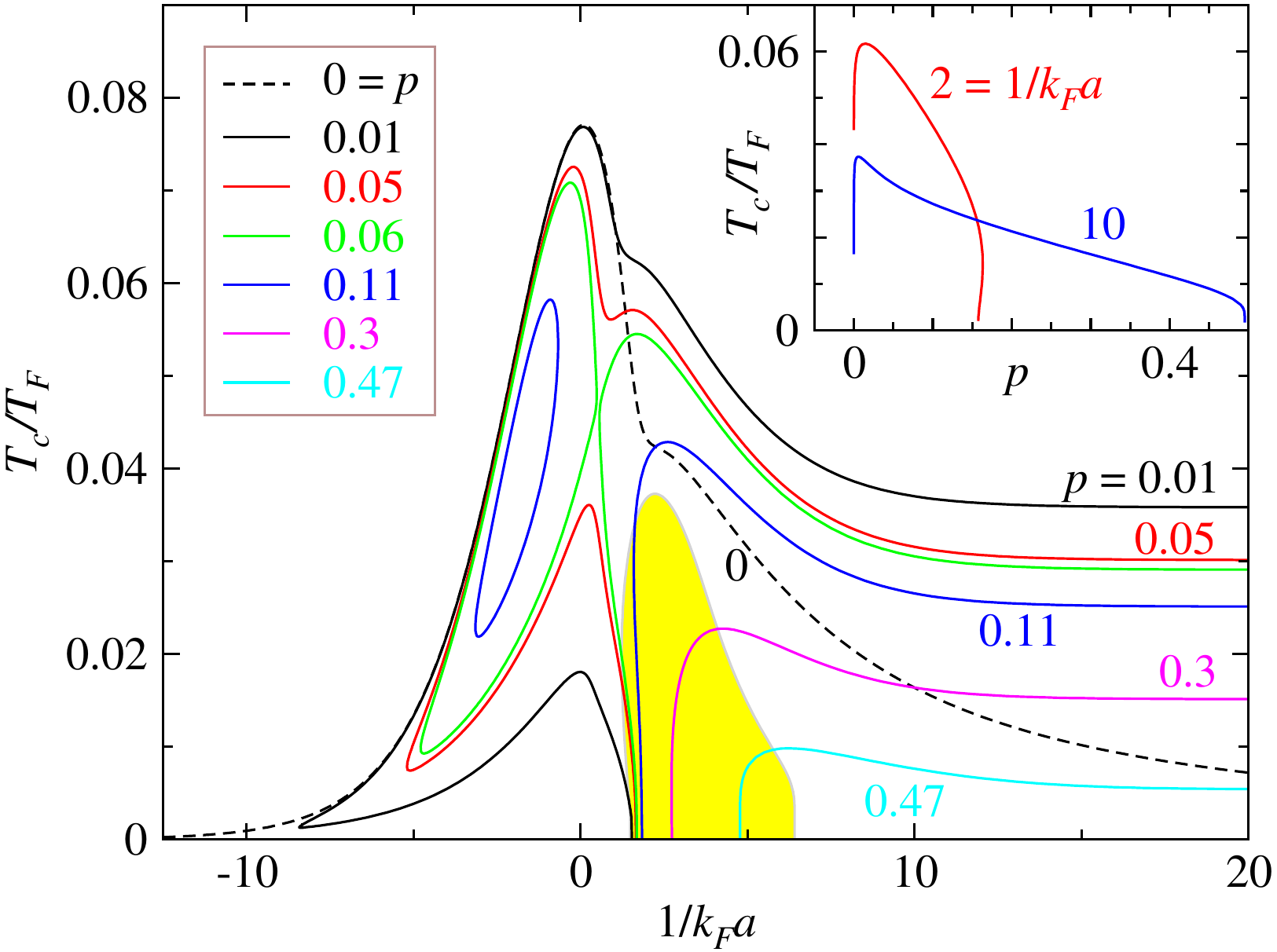}
  \caption{Behavior of $T_c$ versus $1/k_Fa$ with $t/E_F=0.2$ and
    $k_Fd = 0.5$ for varying $p$, as labeled. $T_c$ is unstable in the
    yellow shaded region. Also plotted for comparison is the balanced
    $T_c$ curve (dashed line). Shown in the inset is $T_c$ versus $p$
    for $1/k_Fa = 2$ (red) and 10 (blue lines), manifesting
    enhancement of $T_c$ by imbalance.}
  \label{fig:Tc}
\end{figure}

Shown in Fig.~\ref{fig:Tc} is $T_c$ versus $1/k_Fa$ for varying
imbalance $p$, as labeled, for a physically accessible case with
$t/E_F=0.2$ and $k_Fd = 0.5$. The pSF phase exists only for
$p\lesssim 0.5$. The $T_c$ curves turn back on the BCS side, leading
to an upper and lower $T_c$ for a given $1/k_Fa$, exhibiting typical
ITSF behavior in the BCS and unitary regimes, similar to the 3D
homogeneous case \cite{Chien06,Stability}. Below the lower $T_c$,
phase separation, FFLO and/or PDW states may occur. The $T_c$ solution
in the yellow shaded region is unstable, corresponding to ITSF in the
brown shaded area in Fig.~\ref{fig:T0phase}. Besides the unpaired
normal phase to the left, a pseudogap phase exists above (the upper)
$T_c$.
As $1/k_Fa$ increases, $T_c$ approaches a constant BEC asymptote. This
should be contrasted with the dashed $p=0$ curve, for which $T_c $
decreases with $1/k_Fa$ following Eq.~(\ref{eq:p0Tc}). Therefore,
relative to the $p=0$ case, imbalance may substantially raise $T_c$ on
the BEC side of unitarity. The plot of $T_c$ versus $p$ in the inset  for
$1/k_Fa =2$ and $10$ shows an enhancement for
$p\lesssim 0.1$ and $p\lesssim 0.3$, respectively.

Our calculations reveal that as $d$ increases, the $T=0$ pSF phase shrinks
quickly. For $k_Fd=1$, the upper pSF phase boundary (blue
curve) in Fig.~\ref{fig:T0phase} moves down to $p\sim 0.11$. And for
$d=2$, the pSF phase disappears completely.
The grey shaded ITSF phase extends to $1/k_Fa = +\infty$ at low $p$,
with both an upper and lower $T_c$. Similar reduction of the
superfluid phase can be achieved by decreasing $t$.

\begin{figure}
  \includegraphics[clip,width=3.2in]{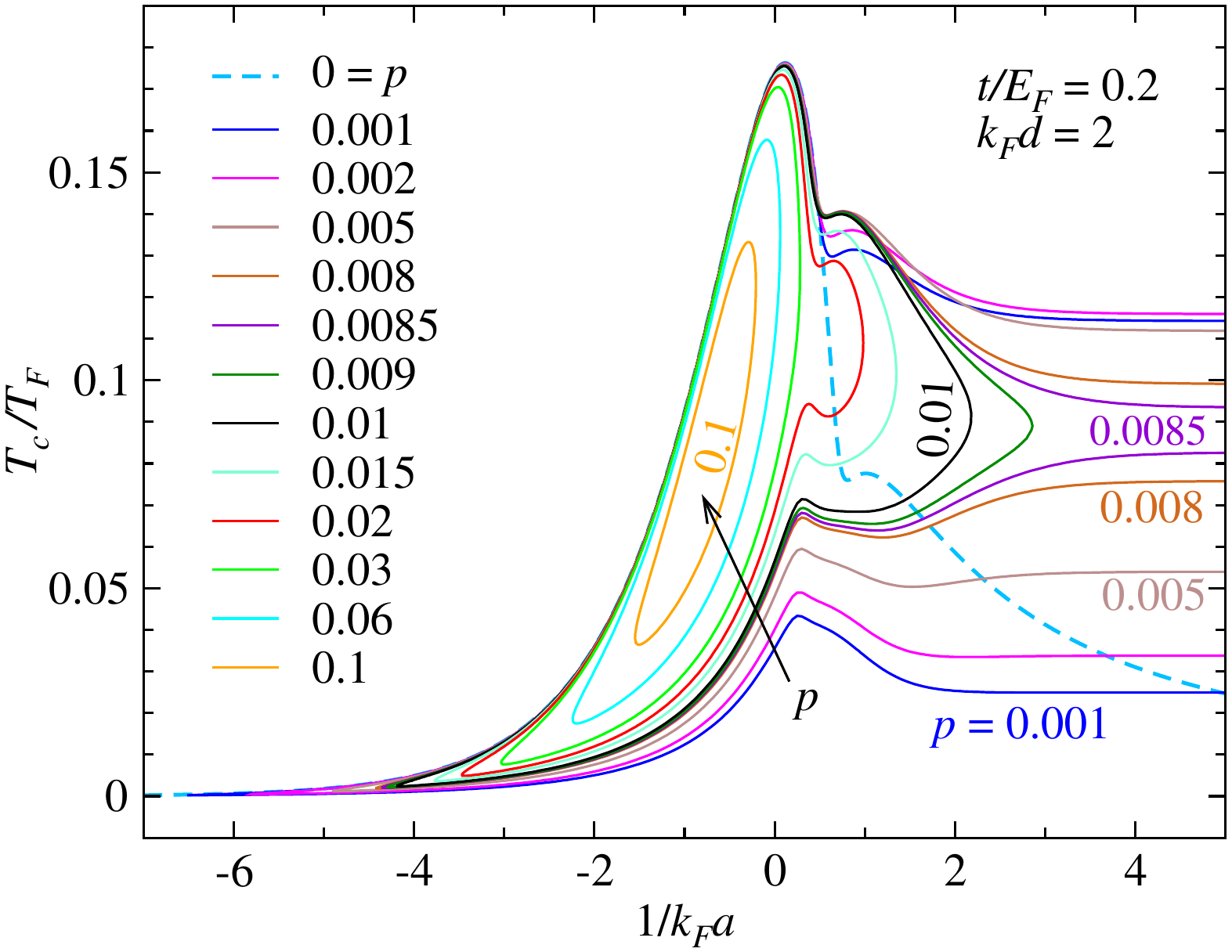}
  \caption{Behavior of $T_c$ versus $1/k_Fa$ with $t/E_F=0.2$ and
    $k_Fd = 2$ for varying $p$, as labeled. Also plotted for
    comparison is the balanced $T_c$ curve (dashed line). The arrow
    points to the $p$ increasing direction.}
  \label{fig:Tc2}
\end{figure}

Shown in Fig.~\ref{fig:Tc2} is a plot of $T_c$ similar to
Fig.~\ref{fig:Tc} except now with increased lattice constant,
$k_Fd = 2$.  The superfluid phase exists only for relatively low $p$,
exhibiting typical ITSF, with the lower $T_c$ extending to
$k_Fa = +\infty$ for $p \lesssim 0.0085$; both the upper and lower
$T_c$'s approach a constant BEC asymptote for these low $p$. For
larger $p \gtrsim 0.009$, there is no superfluid in the deep BEC
regime.  The superfluid phase shrinks to zero
as $p$ increases beyond about 0.135 at $1/k_Fa \approx -0.7$.

\begin{figure}
  \includegraphics[clip,width=3.4in]{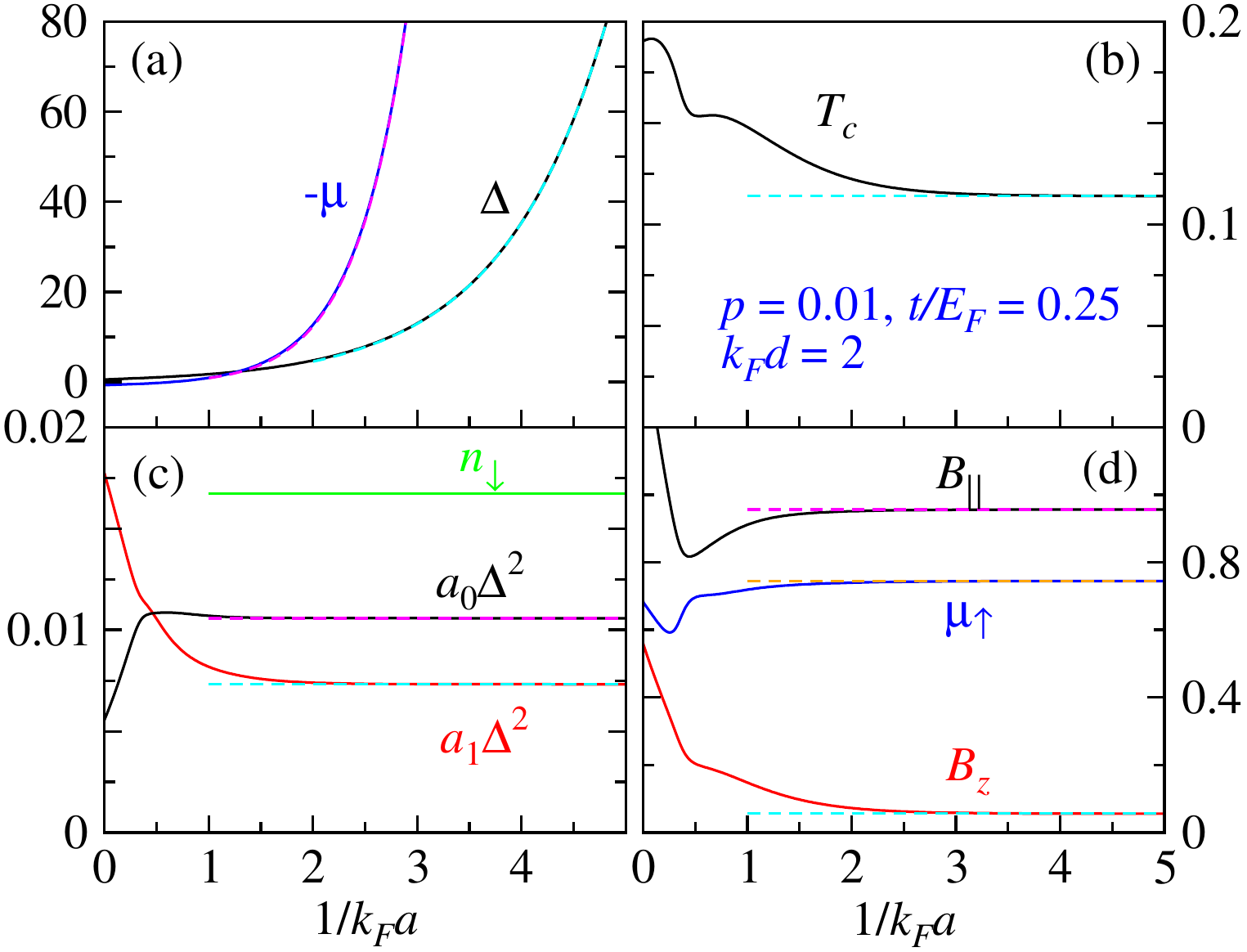}
  \caption{Behavior of (a) $\mu$, $\Delta$, and (b) $T_c$,
    coefficients (c) $a_0$, $a_1$, (d) $B_\parallel$, $B_z$ and
    $\mu_\uparrow$, as a function of $1/k_Fa$ for $p=0.01$,
    $t/E_F=0.25$ and $k_Fd=2$. For comparison, also plotted are the
    BEC asymptotes (dashed lines), as well as $n_\downarrow$. The
    energy unit is $E_F$, and $2m=1$.}
  \label{fig:BEC}
\end{figure}

Finally, we examine in Fig.~\ref{fig:BEC} the asymptotic behavior of
various quantities versus $1/k_Fa$ at $T_c$ in the BEC regime for
$t/E_F =0.25$, $k_Fd=2$ and $p=0.01$. The solid and dashed lines
represent the fully numerical and the BEC asymptotic solutions,
respectively. Figure \ref{fig:BEC}(a) demonstrates that the asymptotic
solutions for $\mu$ and (pseudo)gap $\Delta$ given by
Eqs.~(\ref{eq:muBECp}) and (\ref{eq:gapBECp}) work very well for
$1/k_Fa > 2$. Figure \ref{fig:BEC}(b)-(d) presents $T_c$,
$a_0\Delta^2$, $a_1\Delta^2$, $\mu_\uparrow$, as well as
$B_\parallel$, $B_z = t_Bd^2$ and $\mu_\uparrow$.
They all quickly approach their BEC asymptotes for $1/k_Fa > 3$. In
particular, the constant asymptote for $t_B$ confirms that the
excessive fermion assisted pair hopping dominates in the BEC regime.
As shown in Fig.~\ref{fig:BEC}(c), $ n_p=a_0\Delta^2< n_\downarrow $;
only part of minority fermions form pairs.

We have studied various situations for a big range of $(t,p,d)$ and
found that the superfluidity can be easily destroyed by large $d$ and
small $t$.
Reducing $t$ in Figs.~\ref{fig:Tc} and \ref{fig:Tc2} may shrink the
pSF phase quickly, as shown in Supplementary Fig.~S3.
Overall, in the multidimensional $(t,d,p,T)$ phase space, especially
in the BEC limit, the superfluid phase exists only for small and
intermediate $d$, small $p$, relatively large $t$ and intermediate
(and low) $T$.

To understand the destruction of superfluidity at large $d$ and small
$t$, we note that
when $d$ is large, more fermions will occupy the high $k_\parallel$
states. In addition, a small $t$ may further force the lattice band
fully occupied, so that the Fermi surface becomes
nearly dispersionless as a function of $k_z$.  This makes it extremely
hard to accommodate the excessive majority fermions, which will
necessarily have to occupy high $k_\parallel$ states at a high energy
cost. Furthermore, since $|k_z|\le \pi/d$, the Pauli blocking effect
can no longer be eliminated in the $z$ direction in the BEC
regime, 
so that $t_B$ may be quickly suppressed to zero as $d$ and $p$
increase and/or $t$ decreases at low $T$.

The enhancement and destruction of superfluidity are easily testable
in future experiments. The enhancement also suggests that a small
imbalance is beneficial for achieving superfluidity experimentally.

\begin{acknowledgments}

  This work is supported by NSF of China (Grants No. 11774309 and
  No. 11674283), and NSF of Zhejiang Province of China (Grant No.
  LZ13A040001).
\end{acknowledgments}


\begin{thebibliography}{37}
\bibitem [{not({\natexlab{a}})}]{noteonOL}%
  \BibitemOpen
  \href@noop {} {} ({\natexlab{a}}),\ \bibinfo {note} {an optical lattice in a
  theory paper in the literature often refers to a pure lattice in the context
  of a Hubbard model. Namely, a 1DOL means a simple 1D atomic chain. This is
  different from the 1DOL we study here.}\BibitemShut {Stop}%
\bibitem [{not({\natexlab{b}})}]{noteonEf}%
  \BibitemOpen
  \href@noop {} {}  \bibinfo {note} {Considering changing $t$
  and $d$, here we define $k_F$ anf $T_F$ as given by a homogeneous,
  unpolarized, noninteracting Fermi gas with the same total number density $n$
  in 3D.}\BibitemShut {Stop}%
\bibitem [{not({\natexlab{c}})}]{noteonequations}%
  \BibitemOpen
  \href@noop {} {} \bibinfo {note} {Note that Eq.
  (\ref{eq:PG}) depends on the product $a_0\Delta^2$ and the ratio $a_0/a_1$,
  but not on $\Delta$ separately.}\BibitemShut {Stop}%
\bibitem [{not({\natexlab{d}})}]{noteonphases}%
  \BibitemOpen
  \href@noop {} {} \bibinfo {note} {Note that here we are
  not interested in the pseudogapped normal phase above the upper $T_c$ or
  possible solutions of phase separation and Fulde-Ferrell-Larkin-Ovchinnikov
  states \protect\cite{FF,LO} below the lower $T_c$}\BibitemShut {NoStop}%
\end{thebibliography}
%

\end{document}